\documentstyle[a4 ,12pt]{article}

\input epsf
\begin{document}
\begin{center}
{\bf Implication of the Modified Gottfried Sum Rule} \\
Susumu Koretune\\
Department of Physics, Fukui Medical School\\
Fukui,Matsuoka,910-11 Japan\\
\end{center}
Implication of the modified Gottfried sum rule is discussed by
focusing its theoretical bases,the range of validity ,
and physical interpretation.\\
{\bf (1) Modified Gottfried sum rule }\\ 
Modified Gottfried sum rule takes the following form[1,2]:
\begin{eqnarray}
\lefteqn{\int^1_0\frac{dx}{x}\{F_2^{ep}(x,Q^2)-F_2^{en}(x,Q^2)\}}
\nonumber \\
 &=&\frac{1}{3}\left( 1-\frac{4f_{K}^2}{\pi}\int_{m_Km_N}^{\infty}\frac{d\nu}{\nu^2}
\sqrt{\nu^2-(m_Km_N)^2}\{\sigma^{K^+n}(\nu)-\sigma^{K^+p}(\nu)\}\right) .
\end{eqnarray}
Both-hand sides of the sum rule are related to the same quantity
\begin{equation}
\frac{1}{3\pi}P\int_{-\infty}^{\infty}\frac{d\alpha}{\alpha}
A_3(\alpha ,0).
\end{equation}
Hence the Gottfried sum rule measures the mean $I_3$ of something .
This point is easily understood in the parton model .
The left-hand side of the sum rule multiplied by $3/2$ can be 
expressed in this model as
\begin{eqnarray}
\lefteqn{\int_0^1dx\{\frac{1}{2}u_v - \frac{1}{2}d_v\} +
\int_0^1dx\{\frac{1}{2}\lambda_u - \frac{1}{2}\lambda_d\}
- \int_0^1dx\{- \frac{1}{2}\lambda_{\bar{u}} + 
\frac{1}{2}\lambda_{\bar{d}}\}}\nonumber \\
&=& \frac{1}{2} +\frac{1}{2}\int_0^1dx \{\lambda_u - \lambda_d
+\lambda_{\bar{u}}- \lambda_{\bar{d}}\} .\hspace{6cm}
\end{eqnarray}
This means that the Gottfried sum rule measures the mean $I_3$
of the [(quark) - (anti-quark)] in the proton since 
[(quark) - (anti-quark)] is fixed to be 3 . Therefore we can get
information of the mean quantum number of the sea quarks in 
the proton. \\
Now, because of the O(4) symmetry
in the inclusive lepton-hadron scatterings,
O(4) partial wave expansion is possible in these process.
Then these partial waves were shown to agree with 
the Nachtmann moments apart from trivial kinematic
factors [3].Thus the concept of the moment discussed in the
operator product expansion has a general kinematic meaning.
It can be used in a general context(for details see Ref.[3]and the
papers cited therein) . Then,
all the sum rules discussed here correspond
to the moment at $n=1$ .Since the moment at $n=1$
for the flavor non-singlet had been considered to be well
predicted by the perturbative QCD ,
the $Q^2$ dependence of the Gottfried sum rule was 
checked in Ref.[4], and ,
in the next-to-leading order,it was given roughly 
as $ 0.01[\alpha_s(Q^2)-\alpha_s(Q^2_0)]$. By estimating this
value , the Gottfried sum rule was found to be almost $Q^2$ independent .
The experiment from NMC which reduced the systematic
error greatly was therefore surprising .
The experimental value definitely departs largely from the $1/3$
and it becomes impossible to explain this large departure 
perturbatively. \\
In this situation my formalism sheds new lights [5,6].
In our formalism Gottfried sum rule strongly depends on the pomeron
which appears in the singlet moment at $n=1$ .
If the pomeron is flavor singlet it should be strictly $Q^2$
independent . 
This pomeron reflects the vacuum property of the nucleon[7]
which completely lacks in the perturbative approach .
Thus our approach and the perturbative one is complementary.
The perturbatively predicted $Q^2$ dependence may be shielded
by the large non-perturbative effects or, more practically ,
it can be regarded as negligible compared with the 
non-perturbative contribution . The matching of the 
perturbative $Q^2$ dependence with the non-perturbative one
was done along this line [2].
We considered the followings: In the small $x$ region ,
non-perturbative physics dominates , and it reflects the 
vacuum property of the proton . Some of it appears as the
heavy intrinsic quarks (intrinsic means that it is not the one
produced perturbatively).Because these effects are confined
in the small $x$ region we can see their effects in the 
moments at small $n$ . We consider it the moment at $n=1$ . 
As we go to the larger $x$ region perturbatively produced
pieces become to dominate . Since the contribution
from the small $x$ region is suppressed in the 
moments at large $n$, we consider them as the perturbatively predicted
ones . We took this
moments at large $n$ as the ones above $n=2$. In this way it was shown
that the $Q^2$ dependence in the distribution
can be taken into account in our formalism without conflicting
the success of the perturbative QCD.\\
{\bf (2) Current anti-commutation relations on the null-plane }\\
Our method is based on the current anti-commutation relations 
on the null-plane[5,6]. Intuitively they are obtained as follows .
We take the scalar current $J_a(x)=
:\phi ^{\dagger }(x)\tau_a\phi (x):$.
A canonical commutation relation of the scalar field is
\begin{equation}
[\phi^{\dagger }(x) , \phi (0)]|_{x^+=0}=i\Delta (x),
\end{equation}
where $\Delta (x)$ at $x^+=0$ is $-\epsilon(x^-)\delta (\vec{x}^{\bot})/4$ .\\ 
A connected matrix element of the current product on the null-plane
between the one particle state is 
\begin{eqnarray}
\lefteqn{<p|J_a(x)J_b(0)|p>_c|_{x^+=0} = <p|i\Delta^{(+)}(x):\phi ^{\dagger
  }(x)\tau_a \tau_b\phi (0):}\nonumber \\
&+& i\Delta^{(+)}(x):\phi ^{\dagger }(0)\tau_b \tau_a\phi (x):+
:\phi ^{\dagger}(x)\tau_a\phi (x)\phi ^{\dagger}(0) \tau_b\phi(0):|p>_c .
\end{eqnarray}
At $x^+=0$ , $x^2=2x^+x^- - \vec{x}^{\bot 2}=- \vec{x}^{\bot 2}\leq 0$,
hence the last term does not contribute to the connected part ,and
we get
\begin{eqnarray}
\lefteqn{<p|\{J_a(x),J_b(0)\}|p>_c|_{x^+=0} } \nonumber \\
&=& \Delta^{(1)}(x)<p|:\phi ^{\dagger }(x)\tau_a \tau_b\phi (0): 
+ :\phi ^{\dagger }(0)\tau_b \tau_a\phi (x):|p>_c .  \\
\lefteqn{<p|[J_a(x),J_b(0)]|p>_c|_{x^+=0} }  \nonumber \\
&=& i\Delta (x)<p|:\phi ^{\dagger }(x)\tau_a \tau_b\phi (0): 
+ :\phi ^{\dagger }(0)\tau_b \tau_a\phi (x):|p>_c . 
\end{eqnarray}
Thus by changing $i\Delta (x) \to \Delta^{(1)}(x)$ we can get 
the current anti-commutation relation on the null-plane . 
Now the last term in the left-hand side of Eq.(5)
can not be discarded in the time-like region , hence
the assumption to discard it together with the problem of
the Schwinger terms are obscure in this method .
Thus to justify the intuitive method we need more formal
and sound discussions .
For this purpose I found the Deser-Gilbert-Sudarshan(DGS)
representation [8]which incorporates both the causality and
the spectral condition is an appropriate one .
In this way I found in the flavor $SU(3)\times SU(3)$ model[5,6]
\begin{eqnarray}
\lefteqn{<p|\{J_a^+(x),J_b^+(0)\}|p>_c|_{x^+=0} } \nonumber \\
\lefteqn{= <p|\{J_a^{5+}(x),J_b^{5+}(0)\}|p>_c|_{x^+=0} } \nonumber \\
&=&\frac{1}{\pi }P(\frac{1}{x^-})\delta^2(\vec{x}^{\bot })
[d_{abc}A_c(p\cdot x , x^2=0) + f_{abc}S_c(p\cdot x , x^2=0)]p^+ . 
\end{eqnarray}
If we take the current as
$J_a^{\mu}(x)=\bar{q}(x)\gamma^{\mu}\frac{\lambda_a}{2}q(x)$ and
$J_a^{5\mu}(x)=\bar{q}(x)\gamma^{\mu}\gamma^5\frac{\lambda_a}{2}q(x)$ ,
and if we use the intuitive method , $A_c$ and $S_c$ are related 
to the bilocal current as
\begin{equation} 
S_a^{\mu}(x|0)=
\frac{1}{2}[\bar{q}(x)\gamma^{\mu}\frac{\lambda_a}{2}q(0)
+ \bar{q}(0)\gamma^{\mu}\frac{\lambda_a}{2}q(x)] ,
\end{equation}
\begin{equation}
A_a^{\mu}(x|0)=
\frac{1}{2i}[\bar{q}(x)\gamma^{\mu}\frac{\lambda_a}{2}q(0)
- \bar{q}(0)\gamma^{\mu}\frac{\lambda_a}{2}q(x)] ,
\end{equation}
\begin{equation}
<p|S_a^{\mu}(x|0)|p>_c=p^{\mu}S_a(p\cdot x , x^2) +
x^{\mu}\bar{S}_a(p\cdot x , x^2) ,
\end{equation}
\begin{equation}
<p|A_a^{\mu}(x|0)|p>_c=p^{\mu}A_a(p\cdot x , x^2) +
x^{\mu}\bar{A}_a(p\cdot x , x^2) .
\end{equation}
It should be noted that the causality and the spectral
condition are essential to reach Eq.(8). Further ,
the regular bilocal currents
do not exist in QCD because the each coefficient of the
expansion of them gets the singular piece due to the anomalous
dimension . In other words the
bilocal currents presented here are the singular one .
Combined with the discussions in (1), we can say that 
the bilocal currents can be used as the fixed-mass sum rule
discussed here corresponding to the moment at $n=1$
or as the moment sum rules in the perturbative QCD above $n=2$.\\  
{\bf (3) The sum rules for the mean quantum number of the light sea quarks} \\
From Ref.[2] it is straight-forward to get the hypercharge sum rule as[9]
\begin{eqnarray}
\lefteqn{\frac{1}{2\pi}\frac{2\sqrt{3}}{3}P\int_{-\infty}^{\infty}\frac{d\alpha}{\alpha}
A_8(\alpha ,0)} \nonumber \\
&=&\int_0^1\frac{dx}{x}\{F_2^{\bar{\nu}p}(x,Q^2) +
F_2^{\nu p}(x,Q^2) - 3F_2^{ep}(x,Q^2) - 3F_2^{en}(x,Q^2)\} ,
\end{eqnarray}
\begin{equation}
\frac{1}{2\pi}\frac{2\sqrt{3}}{3}P\int_{-\infty}^{\infty}\frac{d\alpha}{\alpha}
A_8(\alpha ,0)=\frac{1}{3}[2I_{\pi} - I_K^p - I_K^n] , 
\end{equation}
where explicit form of $I_{\pi}$ , $I_K^p$ ,  
and $I_K^n$ and their values are given in Ref.[2]. In principle
they can be determined experimentally . I found that
$I_{\pi}\sim 5.17$ , $I_K^p\sim 2.39$ ,  
and $I_K^n\sim 1.61 $. Thus we obtain 
\begin{equation}
\frac{1}{3}[2I_{\pi} - I_K^p - I_K^n]\sim 2.12 . 
\end{equation}  
Subtracting the contribution from the valence quarks from this ,  we find that 
the mean hypercharge of the light sea quarks is 
$(2.12-1)/2\sim 0.56$ . 
Similarly the mean $I_3$ of the light sea quarks given by the modified 
Gottfried sum rule is $3(0.26-0.33)/4\sim -0.053$ . 
Thus we get the sum rule 
of the mean charge of the light sea quarks in the proton and its value as
\begin{eqnarray}
\lefteqn{<Q>_{light\;sea\;quarks}^{proton}} \nonumber \\
&=&\frac{1}{2}[\{\frac{1}{2\pi}P
\int_{-\infty}^{\infty}
\frac{d\alpha}{\alpha}A_3(\alpha ,0)-\frac{1}{2}\} + \frac{1}{2}\{
\frac{1}{2\pi}\frac{2\sqrt{3}}{3}P\int_{-\infty}^{\infty}\frac{d\alpha}{\alpha}
A_8(\alpha ,0) -1 \} ]  \hspace{5mm} \nonumber \\
&=&\frac{1}{6}(I_{\pi} + I_K^p - 2I_K^n)-\frac{1}{2} \sim 0.2 .
\end{eqnarray}
Similarly for the neutron
\begin{eqnarray}
\lefteqn{<Q>_{light\;sea\;quarks}^{neutron}}\nonumber \\
&=&\frac{1}{2}[-\{\frac{1}{2\pi}P
\int_{-\infty}^{\infty}
\frac{d\alpha}{\alpha}A_3(\alpha ,0)-\frac{1}{2}\} + \frac{1}{2}\{
\frac{1}{2\pi}\frac{2\sqrt{3}}{3}P\int_{-\infty}^{\infty}\frac{d\alpha}{\alpha}
A_8(\alpha ,0) -1 \}]\nonumber \\
&=&\frac{1}{6}(I_{\pi} - 2I_K^p + I_K^n) \sim 0.3 .
\end{eqnarray}
{\bf (4) Conclusion }\\
The modified Gottfried sum rule and its relatives give us
information of the mean quantum number of the sea quarks in the
nucleon . Here we showed that the mean charge of the light sea quarks
in the nucleon can be obtained model independently in
principle. Though the results has been interpreted in the parton model
, they do not depend on this model . They give us the model
independent constraint on the integral of the matrix elements 
of the bilocal currents corresponding to the moment at $n=1$
in an arbitrary frame of the nucleon .\\
{\bf References }\newline
[1]S.Koretune , Phys.Rev.D47(1993)2690.\hspace{0.3cm}
[2]S.Koretune , Phys.Rev.D52(1995)44.\newline
[3]S.Koretune , Prog.Theor.Phys.V.69(1983)1746;V.70(1983)1170E.\newline
[4]D.A.Ross and C.T.Sachrajda ,  Nucl.Phys.B149(1978)497.\newline
[5]S.Koretune , Prog.Theor.Phys.V.72(1984)821.\newline
[6]S.Koretune , Phys.Rev.D21(1980)820.\hspace{0.3cm}
[7]S.Weinberg ,  {\it Lectures on Elementary Particles and Quantum
  Fields Theory}
(M.I.T.Press,1970),p.283.\hspace{0.3cm}
[8]S.Deser , W.Gilbert , and E.C.G.Sudarshan ,
Phys.Rev.115(1959)731.\hspace{0.3cm}
[9]S.Koretune ,  Fukui Medical School preprint :''Mean charge of the
light sea quarks in the proton''.
\end{document}